\documentclass[12pt]{article}
\usepackage{a4wide,epsfig,psfrag,amsmath,amssymb,scalefnt}
\usepackage[dvipsnames]{xcolor}
\usepackage{amsmath,comment,braket}
\usepackage{tabularx}
\usepackage{enumitem}
\usepackage{placeins}
\usepackage{breqn}
\usepackage{slashed}
\usepackage{subcaption}
\usepackage[numbers,sort&compress]{natbib}
\usepackage{hyperref}

\definecolor{ptvblue}{HTML}{0077BB}
\definecolor{ptvcyan}{HTML}{2895be} %33BBEE originally, but it is a bit light on white
\definecolor{ptvteal}{HTML}{009988}
\definecolor{ptvorange}{HTML}{EE7733}
\definecolor{ptvred}{HTML}{CC3311}
\definecolor{ptvmagenta}{HTML}{EE3377}
\definecolor{ptvgrey}{HTML}{828282} %BBBBBB originally, but it is a bit light on white

\usepackage{listings}

\lstdefinestyle{FORMstyle}{
	backgroundcolor=\color{white},
	commentstyle=\color{ptvgrey},
	keywordstyle={[1]\color{ptvred}},
	keywordstyle={[2]\color{ptvred}}, %glyellow
	keywordstyle={[3]\color{ptvteal}}, %glyellow
	keywordstyle={[4]\color{ptvorange}},
	keywordstyle={[5]\color{ptvblue}},
	keywordstyle={[6]\color{ptvcyan}},
	keywordstyle={[7]\color{ptvteal}},
	stringstyle=\color{ptvteal},
	basicstyle=\color{black}\ttfamily\bfseries\normalsize,
    columns=flexible,
	breakatwhitespace=false,
	breaklines=true,
	captionpos=b,
	keepspaces=true,
	numbers=none,
	numbersep=2pt,
	numberstyle=\color{black}\scriptsize,
	showspaces=false,
	showstringspaces=false,
	showtabs=false,
	tabsize=2,
	frame=lr,framesep=2pt,framerule=0pt
}

\lstdefinelanguage{FORM}{
	sensitive = false,
	comment=[f]{*},
	moredelim=[s][\color{ptvcyan}]{`}{'},
	string=[b]",
	alsoletter={.,\#,\$},
	morekeywords={[1]also,antiputinside,antisymmetrize,argexplode,argimplode,argtoextrasymbol,
		argument,chainin,chainout,chisholm,chop,contract,cyclesymmetrize,denominators,discard,disorder,
		evaluate,factarg,factdollar,frompolynomial,goto,identify,idnew,idold,ifmatch,ifnomatch,inexpression,inside,label,
		makeinteger,many,merge,multi,multiply,normalize,once,only,putinside,ratio,rcyclesymmetrize,
		redefine,renumber,replaceloop,select,setexitflag,shuffle,sort,splitarg,splitfirstarg,
		splitlastarg,strictrounding,stuffle,sum,symmetrize,term,tofloat,topolynomial,torational,tospectator,
        totensor,tovector,trace4,tracen,transform,tryreplace},
	morekeywords={[2]auto,autodeclare,cfunction,cfunctions,cleartable,commuting,commuteinset,
		compress,createspectator,ctable,ctensor,ctensors,deallocatetable,dimension,endmodel,
		extrasymbols,fill,fillexpression,fillindex,function,functions,funpowers,index,indices,
		indexes,insidefirst,load,metric,model,modulus,nfunction,nfunctions,ntable,ntensor,ntensors,
		nwrite,off,on,particle,polyfun,polyratfun,processbucketsize,propercount,save,set,slavepatchsize,
		symbol,symbols,table,tensor,tensors,threadbucketsize,unittrace,vector,vectors,vertex,write},
	morekeywords={[3]copyspectator,gfactorized,globalfactor,globalfactorized,global,lfactorized,
		localfactor,localfactorized,local},
	morekeywords={[4]abracket,antibracket,abrackets,antibrackets,bracket,brackets,factorize,format,
		moduleoption,nfactorize,print,nprint,nunfactorize,unfactorize},
	morekeywords={[5]abs_,agm_,bernoulli_,binom_,conjg_,content_,count_,d_,dd_,delta_,deltap_,denom_,
		distrib_,div_,dum_,dummy_,dummyten_,e_,euler_,eexp_,exp_,exteuclidean_,extrasymbol_,fac_,factorin_,farg_,
		firstbracket_,firstterm_,float_,g5_,g6_,g7_,g_,gamma_,gcd_,gi_,id_,integer_,invfac_,inverse_,makerational_,
		match_,max_,maxpowerof_,min_,minpowerof_,mod_,mod2_,mul_,mzv_,mzvhalf_,nargs_,nterms_,numfactors_,partitions_,
		pattern_,perm_,poly_,prime_,putfirst_,random_,ranperm_,rem_,replace_,reverse_,root_,setfun_,sig_,
		sign_,sizeof_,sum_,sump_,table_,tbl_,term_,termsin_,termsinbracket_,theta_,thetap_,topologies_,
		sqrt_,ln_,sin_,cos_,tan_,asin_,acos_,atan_,atan2_,sinh_,cosh_,tanh_,asinh_,acosh_,atanh_,li2_,
		lin_,coeff_,num_,den_,xarg_,dimension_,factor_,sep_,iarg_,parg_,pi_,ee_,em_,
        diagrams_,topo_,node_,edge_,block_,onepi_},
	morekeywords={[1].sort,.end,.store,.global,.clear},
	morekeywords={[1]collect,delete,drop,emptyspectator,hide,inparallel,intohide,keep,ndrop,nintohide,nhide,
		notinparallel,nskip,nunhide,pophide,pushhide,removespectator,skip,unhide},
	morekeywords={[6]\#add,\#addseparator,\#append,\#appendpath,\#break,\#breakdo,\#call,\#case,\#clearoptimize,
		\#close,\#closedictionary,\#commentchar,\#continuedo,\#create,\#debug,\#default,\#define,\#do,\#else,\#elseif,\#enddo,
		\#endfloat,\#endif,\#endinside,\#endprocedure,\#endswitch,\#exchange,\#external,\#factdollar,\#fromexternal,\#if,
		\#ifdef,\#ifndef,\#include,\#inside,\#message,\#opendictionary,\#optimize,\#pipe,\#preout,\#prependpath,
		\#printtimes,\#procedure,\#procedureextension,\#prompt,\#redefine,\#remove,\#reset,\#reverseinclude,
		\#rmexternal,\#rmseparator,\#setexternal,\#setexternalattr,\#setrandom,\#show,\#skipextrasymbols,\#sortreallocate,
		\#startfloat,\#switch,\#system,\#terminate,\#timeoutafter,\#toexternal,\#undefine,\#usedictionary,\#write,\#},
	morekeywords={[1]break,case,default,do,else,elseif,enddo,endif,endrepeat,endswitch,endwhile,if,repeat,switch,while},
	morekeywords={[7]count,match,expression,occurs,findloop,multipleof,coefficient},
    morekeywords={[1]tablebase,open,readonly,use,testuse,apply,load},
    literate=
    *{\$tmp}{{{\color{ptvcyan}\$tmp}}}4
    {\$prcLocalDollar}{{{\color{ptvcyan}\$prcLocalDollar}}}{15}
    {n?}{{{\color{ptvmagenta}n?}}}2
    {f?}{{{\color{ptvmagenta}f?}}}2
    {?a}{{{\color{ptvmagenta}?a}}}2
}

\newcommand{\FORMfi}{\textsc{Form}~5}
\newcommand{\FORM}{\textsc{Form}}
\newcommand{\TFORM}{\textsc{TForm}}
\newcommand{\PARFORM}{\textsc{ParForm}}
\newcommand{\FLINT}{\textsc{Flint}}

\newcommand{\forminline}[1]{\lstinline[language=FORM,style=FORMstyle]{#1}}
\lstnewenvironment{formblock}
    {\lstset{language=FORM,style=FORMstyle,
    }}
    {\vspace{-\parskip}} % hack to remove extra parskip after end{lstlisting}
\lstnewenvironment{formblocknohl}
    {\lstset{style=FORMstyle,
    }}
    {\vspace{-\parskip}} % hack to remove extra parskip after end{lstlisting}
\lstnewenvironment{formblocksmall}
    {\lstset{language=FORM,style=FORMstyle,
    basicstyle=\color{black}\ttfamily\bfseries\small,
    }}
    {\vspace{-\parskip}} % hack to remove extra parskip after end{lstlisting}
\lstnewenvironment{formblocksmallnohl}
    {\lstset{style=FORMstyle,
    basicstyle=\color{black}\ttfamily\bfseries\small,
    }}
    {\vspace{-\parskip}} % hack to remove extra parskip after end{lstlisting}

\title{FORM Version 5.0}
\author{
    J.~Davies$^{a}$,
    T.~Kaneko$^{b}$,
    C.~Marinissen$^{c}$,
    T.~Ueda$^{d}$,
    J.A.M.~Vermaseren$^{c}$
    \\[1mm]
    {\small\textit{(a) Department of Mathematical Sciences, University of
    Liverpool,}}
    \\
    {\small\textit{Liverpool, L69 3BX, UK}}
    \\
    {\small\textit{(b) High Energy Accelerator Research Organization (KEK),
1-1 Oho, Tsukuba,}}
    \\
    {\small\textit{Ibaraki 305-0801, Japan}}
    \\
    {\small\textit{(c) Nikhef Theory Group, Science Park 105, 1098 XG Amsterdam, The Netherlands}}
    \\
    {\small\textit{(d) Department of Mathematics, Faculty of Medicine, Juntendo University,}}
    \\
    {\small\textit{Chiba 270-1695, Japan}}
}

\date{January 2026}

\begin{document}
\maketitle
\thispagestyle{empty}

\begin{abstract}
We present \FORMfi{}, a major release of the symbolic-manipulation system \FORM.
Version~5 introduces an integrated diagram generator, based on the \textsc{Grace} graph-generator, to produce Feynman diagrams directly from \FORM{} scripts. This release also adds support for arbitrary precision floating point coefficients, together with statements for the numerical evaluation of common mathematical functions as well as multiple zeta values and Euler sums. In addition, \FORMfi{} provides an interface to the \FLINT{} library, offering substantially faster polynomial arithmetic.
Various further functions and commands have been added alongside these major features, as well as performance improvements for \TFORM{} and improved compression of \FORM's temporary files.
Compatibility with the previous release, \FORM~4.3.1, is retained except where prior behaviour contradicted the manual or was experimental.
\end{abstract}

\thispagestyle{empty}
\newpage

\setcounter{tocdepth}{2}
\tableofcontents
\newpage

%%%%%%%%%%%%%%%%%%%%%%%%%%%%%%%%%%%%%%%%%%%%%%%%%%%%%%%%%%%%%%%%%%%%%%
\section{Introduction}
We would like to begin by prominently acknowledging the following colleagues for code contributions to the \FORM{} project over the last few years:
R.~L.~Delgado,
F.~Herren,
S.~P.~Jones,
F.~Lorkowski,
V.~Magerya,
A.~Maier,
and
M.~Reininghaus.
The development and maintenance of software packages which are key to the research outputs
of the theoretical physics community is of paramount importance, and should be considered to be
equally as impactful as the results which they enable.

Over the past decades the symbolic manipulation system \FORM{}~\cite{Vermaseren:2000nd} has continued to evolve in response to the increasing complexity of problems encountered in e.g.~high-energy physics, number theory and large-scale algebraic computations in general. Each new release has extended \FORM{}~\cite{Tentyukov:2007mu,Kuipers:2012rf,Kuipers:2013pba,Ruijl:2017dtg,Ueda:2020wqk} in ways that enabled users to push further into domains requiring ever larger expressions, more sophisticated pattern matching, and more efficient polynomial operations. Recent user demands still motivate the further development of \FORM; it has a very active user base and packages are still developed for \FORM{} or which use \FORM{} internally, such as
ANATAR~\cite{Duhr:2025oil},
GoSam-3.0~\cite{Braun:2025afl},
HepLib~\cite{Feng:2021kha},
HyperFORM~\cite{Kardos:2025klp},
IBIS~\cite{vanHoegaerden:2025nmh},
LoopScalla~\cite{Shtabovenko:2025lxq},
MaRTIn~\cite{Brod:2024zaz},
MultivariateApart~\cite{Heller:2021qkz},
OPITeR~\cite{Goode:2024cfy}
and pySecDec~\cite{Heinrich:2023til}.

In addition to 50 bug fixes of existing features and some performance improvements\footnote{In particular, the parallel scaling of \TFORM{} has improved substantially for some benchmarks.} (which are fully described in the \FORMfi{} release notes \url{https://github.com/form-dev/form/wiki/Release-Notes-FORM-5.0.0}), the main new features of \FORMfi{} include:
\begin{itemize}
    \item a built-in diagram generator based on the \textsc{Grace} system~\cite{Kaneko:1994fd}, allowing the user to generate scattering amplitudes directly in a \FORM{} script without relying on external tools,
    \item support for arbitrary precision floating-point term coefficients, and the numerical evaluation of a variety of functions including multiple zeta values and Euler sums, and
    \item an interface to \FLINT{}~\cite{flint}, providing greatly improved performance for polynomial arithmetic.
\end{itemize}
The test suite has also been improved significantly, though further work is
required for it to have sufficient coverage to make future development more
robust.

This paper describes the new features and changes in detail. In Section~\ref{sec:smallChanges}, we discuss some recent bug fixes and small changes made to the \FORM{} scripting language. We then discuss in Section~\ref{sec:features} the main new features of \FORMfi.  
In Section~\ref{sec:deprecations} we discuss plans to remove some features in the future which, to our knowledge, are not used and create a substantial maintenance burden; these features exist in \FORMfi{}, but with a ``deprecated'' status.
We conclude with a short outlook and an invitation to the community to contribute to the ongoing development of the \FORM{} project.

\FORM's development repository is hosted at: \url{https://github.com/form-dev/form}. See the included \texttt{INSTALL} file for up-to-date installation instructions, particularly regarding the new library dependencies: \FLINT{}, MPFR and Zstandard.
The user manual for \FORMfi{} can be found here:
\url{https://form-dev.github.io/form-docs/stable/manual/},
and a version of the manual which continuously reflects the current state of the
development branch can be found here:
\url{https://form-dev.github.io/form-docs/master/manual/}.

%%%%%%%%%%%%%%%%%%%%%%%%%%%%%%%%%%%%%%%%%%%%%%%%%%%%%%%%%%%%%%%%%%%%%%
\section{Bug Fixes and Small Changes}
\label{sec:smallChanges}
In this section we describe fixes or small changes which may impact the user, either in terms of
\FORM{} scripts or scripting around a \FORM-based workflow. \FORMfi{} does not break compatibility
with \FORM{} 4.3.1 scripts except in cases where behaviour differed from descriptions in the manual
or the removal of ``experimental'' functions such as \forminline{topologies\_} (see Section~\ref{sec:diagen}).

%%%%-----------------------------------------------------------------
\subsection{Changes between Version 4.2 and 4.3}
Various changes were made and additional features added between \FORM{} Version 4.2.1 and 4.3.1, which were
never described in a paper. Summaries of these changes can be found on the following pages:
\begin{itemize}
    \item \url{https://github.com/form-dev/form/wiki/Release-Notes-Form-4.2.1}
    \item \url{https://github.com/form-dev/form/wiki/Release-Notes-FORM-4.3.0}
    \item \url{https://github.com/form-dev/form/wiki/Release-Notes-FORM-4.3.1}
\end{itemize}

%%%%-----------------------------------------------------------------
\subsection{Fixed \texttt{IntoHide} and the addition of \texttt{NIntoHide}}
The specification \forminline{IntoHide;} now correctly hides all active expressions at the end of the
current module. \forminline{NIntoHide} has been added to allow negation of \forminline{IntoHide} status,
such that \forminline{IntoHide}/\forminline{NIntoHide} can
be used in the same way as \forminline{Drop}/\forminline{NDrop} and \forminline{Hide}/\forminline{NHide}.

%%%%-----------------------------------------------------------------
\subsection{Forced ``Multi-run'' Mode}
In an effort to reduce crashes when running concurrent \FORM{} jobs with a common \texttt{FORMTMP} temporary
directory for scratch and sort files, the ``multi-run'' mode (which includes the process identification number
in the temporary file names) is now always enabled. The command-line option \texttt{-M} is ignored.

%%%%-----------------------------------------------------------------
\subsection{Output Modes}
\subsubsection{Fortran}
When using the Fortran output modes (\forminline{Format Fortran;}, \forminline{Format DoubleFortran;}, \forminline{Format QuadFortran;}),
the floating-point suffix (``\texttt{.}'', ``\texttt{.D0}'', ``\texttt{.Q0}'') is now correctly included for integer literals
larger than $2^{31}-1$, avoiding errors from Fortran compilers.

\subsubsection{Mathematica}
To improve compatibility, \forminline{Format Mathematica;} now encloses dot products and entire
expressions in parentheses. This helps when importing multi-line expressions which contain empty
lines into Mathematica, or expressions which contain powers of scalar products.
Thus, the following example produces a result which is correctly parsed by Mathematica.
\begin{minipage}[t]{0.45\textwidth}
\begin{formblock}
Vector p,q;
Symbol x;

Local F = (1 + x) * (1 + p.q)^2;

Format Mathematica;
Bracket x;
Print +s;
.end
\end{formblock}
\end{minipage}\hfill
\begin{minipage}[t]{0.45\textwidth}
\begin{formblocknohl}
F = (

    + x * (
       + 1
       + 2*(p.q)
       + (p.q)^2
       )

    + 1
       + 2*(p.q)
       + (p.q)^2
      );
\end{formblocknohl}
\end{minipage}

%%%%-----------------------------------------------------------------
\subsection{Statistics Printing}
The sorting statistics are now optionally printed with additional ``human readable'' numbers when enabled with \forminline{On HumanStats;}.
Thus the statistics may, for example, have the form:
\begin{formblock}
Time =      10.01 sec    Generated terms = 1234567890 (  1 B  )
            test         Terms in output =       1234 (  1 K  )
                         Bytes used     =123456789000 (115 GiB)
\end{formblock}

%%%%-----------------------------------------------------------------
\subsection{New Warnings}
\subsubsection{Missing or conflicting dollar variable ModuleOptions}
\label{sec:dollarwarnings}
In \TFORM{} and \PARFORM, the use of dollar variables without an appropriate accompanying \forminline{ModuleOption}
specifying them as \texttt{local}, \texttt{maximum}, \texttt{minimum} or \texttt{sum} now prints a warning by default, in an
effort to avoid modules running in sequential mode without the user realising. For example, the following \FORM{}
script:
\begin{formblock}
Symbol x,n;
Local test = x;
Identify x^n?$tmp = n;
Print;
.end
\end{formblock}
will print a warning when running with \TFORM{}:
\begin{formblocknohl}
Warning: This module is forced to run in sequential mode due to $-variable: $tmp
\end{formblocknohl}
Previously, this warning was only printed when using the non-default option
\begin{formblock}
On AllWarnings;
\end{formblock}

Additionally, a warning is now printed in case conflicting \forminline{ModuleOption}
specifications are provided within the same module. For example, the following
script:
\begin{formblock}
$tmp = 1;
ModuleOption local $tmp;
ModuleOption sum $tmp;
.end
\end{formblock}
will print
\begin{formblocknohl}
Warning: Conflicting module options for $-variable; later option ignored: $tmp
\end{formblocknohl}

\subsubsection{Save files and expression name length}
Attempting to \forminline{Save} stored expressions with names longer than 16 characters now
triggers a warning, for example:
\begin{formblocknohl}
Warning: saved expr name over 16 char: abcdefghijklmnopq
\end{formblocknohl}
While names up to 19 characters work in principle, this behaviour should not be relied upon.
Saving expressions with names of 20 characters or more does not work correctly.
The manual has long stated the 16-character limit for expression name length, though it has
never been enforced. To maintain compatibility with existing scripts it is still not enforced,
but this new warning draws attention to the issue.

%%%%-----------------------------------------------------------------
\subsection{Free placement of dollar variable ModuleOption statements}
Related to Section~\ref{sec:dollarwarnings}, \forminline{ModuleOption} statements pertaining to the parallel
treatment of dollar variables may now be specified anywhere after the dollar variable's definition, rather than
strictly at the end of the module. As well as making the parallel behaviour of dollar variables more clear,
this is particularly useful when writing procedures, perhaps as part of a library, which use dollar variables
but do not contain a \forminline{.sort}. The following structure is now valid:

\begin{minipage}{\textwidth} % force no page breaks in the listing
\begin{formblock}
#Procedure sortlessProcedure
$prcLocalDollar = count_(x,1);
ModuleOption local $prcLocalDollar;
* more executable statements here
#EndProcedure

#Call sortlessProcedure
* more executable statements here
.sort
\end{formblock}
\end{minipage}

%%%%-----------------------------------------------------------------
\subsection{\#ContinueDo}
The \forminline{\#ContinueDo} pre-processor statement may be used to skip to the next iteration of a pre-processor \forminline{\#Do} loop.
The following example yields \forminline{f(1,2,4,5)}:

\begin{minipage}{\textwidth} % force no page breaks in the listing
\begin{formblock}
NFunction f;
Local F = 1;
#Do i = 1,5
   #If `i' == 3
      #ContinueDo
   #EndIf
   Multiply right f(`i');
#EndDo
ChainIn f;
Print;
.end
\end{formblock}
\end{minipage}

Additionally, \forminline{\#ContinueDo <n>} may be used in the case of nested \forminline{\#Do} loops to skip to the next iteration of the \forminline{\#Do} loop which is \forminline{n} levels higher. The following example yields \forminline{i*j*k + i*j*k^2 + i^2*j*k + i^2*j*k^2 + i^3*j*k + i^3*j*k^2}\,:

\begin{minipage}{\textwidth} % force no page breaks in the listing
\begin{formblock}
Symbol i,j,k;
Local F =
#Do i = 1,3
   #Do j = 1,3
      #Do k = 1,3
         #If `k' == 3
            #ContinueDo 3
         #EndIf
         + i^`i'*j^`j'*k^`k'
      #EndDo
   #EndDo
#EndDo
;
Print;
.end
\end{formblock}
\end{minipage}

%%%%-----------------------------------------------------------------
\subsection{Version information}
A new command-line switch ``\texttt{-vv}'' has been added, which prints
a list of the enabled and disabled features in the compilation of \FORM,
as well as the versions of linked libraries used at runtime. Thus, running
\texttt{form -vv} will yield output similar to:
\begin{formblocknohl}
FORM 5.0.0 (Jan 27 2026, v5.0.0)
-backtrace  +flint=3.4.0  +gmp=6.3.0   -mpi    -pthreads  +zlib=1.3.1
+debugging  +float        +mpfr=4.2.2  +posix  -windows   +zstd=1.5.7
Compiler: GCC 15.2.0
Architecture: arm64
\end{formblocknohl}
This information should be included when raising a bug report against \FORM.

Additionally, a new pre-processor variable \forminline{`SubSubVersion_'} has been
added. Alongside \forminline{`Version_'} and \forminline{`SubVersion_'} this allows
for finer-grained detection of whether certain features or bug fixes are present
in \FORM, for example in library procedures which depend on them.

%%%%%%%%%%%%%%%%%%%%%%%%%%%%%%%%%%%%%%%%%%%%%%%%%%%%%%%%%%%%%%%%%%%%%%
\section{New Features}
\label{sec:features}
In this section we list the larger new features which form the basis of \FORMfi.

%%%%-----------------------------------------------------------------
\subsection{Built-in Diagram Generator}
\label{sec:diagen}
\FORMfi{} provides an interface to the graph generator of the \textsc{Grace} system~\cite{Kaneko:1994fd}, which has been
reprogrammed as an easy-to-understand \texttt{C++} library for this purpose by T.~Kaneko;
its code is included in the \FORM{} distribution and is licensed under
GPLv3.
Syntax has been introduced to the \FORM{} language to define models and generate graphs directly within
\FORM{} scripts, which is outlined below. \FORM~4.2.1 featured the experimental \forminline{topologies\_} function, which has now been removed.

To begin, one must define a \forminline{Model} which contains the fields and vertices to be used for the generation of Feynman graphs.
A \forminline{Model} definition is finished with \forminline{EndModel;}.
A \forminline{Particle} is defined with the syntax:
\begin{formblock}
Particle particlename[,antiparticlename] [,<sign><spin>] [,external];
\end{formblock}
A \forminline{Particle} may optionally have an antiparticle name assigned to it;
if no antiparticle is specified the particle is its
own antiparticle. The \texttt{<sign>} denotes bosonic (\texttt{+}) or fermionic (\texttt{-}) statistics and the
\texttt{<spin>} denotes the dimension of the particle's spin representation; for example, a scalar is defined with
\texttt{+1}, an electron with \texttt{-2} and a gluon with \texttt{+3}.
The spin information is used by the generator to forbid invalid vertices, but does not ultimately affect the generation
of the graphs.
Neglecting the sign and spin information is equivalent to specifying \texttt{+1}.
The \texttt{external} option specifies that the \forminline{Particle} should appear only as an external leg.

Following the \forminline{Particle} definitions inside the \forminline{Model} scope,
one defines interaction vertices. An N-point \forminline{Vertex} is defined
as follows, where N $\geq 2$:
\begin{formblock}
Vertex particle1,...,particleN: coupling;
\end{formblock}
The coupling should be a (product of) symbol(s) to integer powers.
Symbols used here will be declared automatically if not already declared.

Multiple models, with different names, may be defined within a single \FORM{} script.
Some concrete examples follow:

\begin{minipage}{0.45\textwidth}
\begin{formblock}
Model PHI3;
   Particle phi, +1;
   Vertex phi,phi,phi: g;
EndModel;

Model PHI4;
   Particle phi, +1;
   Vertex phi,phi,phi,phi: g^2;
EndModel;
\end{formblock}
\end{minipage}\hfill
\begin{minipage}{0.45\textwidth}
\begin{formblock}
Model QCD;
   Particle qua,QUA,  -2;
   Particle gho,GHO,  -1;
   Particle glu,      +3;
   Vertex QUA,qua,glu: g;
   Vertex GHO,gho,glu: g;
   Vertex glu,glu,glu: g;
   Vertex glu,glu,glu,glu: g^2;
EndModel;
\end{formblock}
\end{minipage}

Once a \forminline{Model} has been defined, Feynman graphs may be generated with the \forminline{diagrams\_} function,
which has the following syntax:

\begin{minipage}{\textwidth} % force no page breaks in the listing
\begin{formblock}
diagrams_(model_name, incoming_particle_set, outgoing_particle_set,
          external_momenta_set, internal_momenta_set,
          number_of_loops_or_couplings, options);
\end{formblock}
\end{minipage}

Here, \texttt{model_name} is the name of the user-defined \forminline{Model}, and \texttt{incoming_particle_set} and
\texttt{outgoing_particle_set} are \FORM{} sets of \forminline{Particle} names, for example ``\texttt{\{phi\}}'' or
``\texttt{\{glu,glu\}}'', representing the desired scattering amplitude.
At two or more loops, it is also possible to generate vacuum graphs by specifying empty sets for both the incoming
and outgoing particles.\footnote{Vacuum graphs have been validated
to five loops against \textsc{FeynGraph}
\url{https://jens-braun.github.io/FeynGraph/}.}
\texttt{external_momenta_set} and
\texttt{internal_momenta_set} are \FORM{} sets of vectors to be used to represent internal and external particle
momenta, respectively. These can be defined sets, such as
\begin{formblock}
Vector q1,...,q10, k1,...,k10;
Set ext : q1,...,q10;
Set int : k1,...,k10;
\end{formblock}
or defined dynamically in the call of \forminline{diagrams\_} by providing, for example, \forminline{\{q1,q2\}}.
The vectors in the external and internal sets must not come with a minus sign, and the sets must not contain any
repeated entries.
The \texttt{number_of_loops_or_couplings} can be set either to a number denoting the number of loops required,
or to a specific power of the coupling constants used in the \forminline{Model} vertices, for example
``\forminline{g^2}'' or ``\forminline{gs^2 * gw^2}''.

Finally, the \texttt{options} argument can be used to control the generation and output formatting of the graphs.
The options are \FORM{} pre-processor variables, and should be added together. If no filtering options are specified
(by passing ``\texttt{0}'' or omitting the argument entirely), all connected graphs are generated.
The graph-filtering keywords are defined to be compatible with their counterparts in \textsc{Qgraf}~\cite{Nogueira:1991ex} and
are given below. Options which are inverse to each other may not be specified simultaneously.
The user must be careful not to accidentally specify a keyword twice; this will not have the intended effect, but rather
generate a different keyword entirely. This subtlety may be improved in the future.
\begin{description}[leftmargin=!,labelwidth=\widthof{\textbf{\texttt{\color{ptvcyan}`NoTadpole_'}, \texttt{\color{ptvcyan}`Tadpole_'}:}}]
    \item[\textbf{\texttt{\color{ptvcyan}`OnePI_'}, \texttt{\color{ptvcyan}`OnePR_'}}:] generate only one-particle irreducible (reducible) graphs.
    \item[\textbf{\texttt{\color{ptvcyan}`OnShell_'}, \texttt{\color{ptvcyan}`OffShell_'}:}] generate only graphs without (with) self-energy corrections on external lines.
    \item[\textbf{\texttt{\color{ptvcyan}`NoSigma_'}, \texttt{\color{ptvcyan}`Sigma_'}:}] generate only graphs without (with) any self-energy corrections on any line.
    \item[\textbf{\texttt{\color{ptvcyan}`NoSnail_'}, \texttt{\color{ptvcyan}`Snail_'}:}] generate only graphs without (with) snails.
    \item[\textbf{\texttt{\color{ptvcyan}`NoTadpole_'}, \texttt{\color{ptvcyan}`Tadpole_'}:}] generate only graphs without (with) tadpoles.
    \item[\textbf{\texttt{\color{ptvcyan}`Simple_'}, \texttt{\color{ptvcyan}`NotSimple_'}:}] generate only graphs without (with) any vertices connected by two or more edges.
    \item[\textbf{\texttt{\color{ptvcyan}`Bipart_'}, \texttt{\color{ptvcyan}`NonBipart_'}:}] generate only bipartite (non-bipartite) graphs.
    \item[\textbf{\texttt{\color{ptvcyan}`CyclI_'}, \texttt{\color{ptvcyan}`CyclR_'}:}] generate only cycle irreducible (reducible) graphs.
    \item[\textbf{\texttt{\color{ptvcyan}`Floop_'}, \texttt{\color{ptvcyan}`NotFloop_'}:}] generate only graphs which do not (do) contain closed fermion loops with an odd number of vertices.
\end{description}
Graph generation is further controlled with the options:
\begin{description}[leftmargin=!,labelwidth=\widthof{\textbf{\texttt{\color{ptvcyan}`TopologiesOnly_'}:}}]
    \item[\textbf{\texttt{\color{ptvcyan}`WithSymmetrizeI_'}, \texttt{\color{ptvcyan}`WithSymmetrizeF_'}:}] symmetrize between the initial (final) state particles. For example, when generating a boson propagator one could define both external particles as incoming and provide the \forminline{`WithSymmetrizeI_'} option.
    \item[\textbf{\texttt{\color{ptvcyan}`TopologiesOnly_'}:}] generate only the distinct topologies which appear in the requested amplitude. In this mode, \forminline{Particle} information is not given in the output, but only the momenta flowing into each vertex.
    The topology numbering in the \forminline{topo_} tags is consistent with and without
    this option, such that graphs in the full output are produced with their topologies
    already identified. Preparatory work can be performed efficiently at the level of the 
    topologies before processing the full graph output.
\end{description}
In addition to the filtering keywords, the following options control the formatting of the output:
\begin{description}[leftmargin=!,labelwidth=\widthof{\textbf{\texttt{\color{ptvcyan}`TopologiesOnly_'}:}}]
    \item[\textbf{\texttt{\color{ptvcyan}`WithEdges_'}:}] produce also \forminline{edge_} functions which contain propagator momenta and the numbers of the vertices to which they connect.
    \item[\textbf{\texttt{\color{ptvcyan}`WithoutNodes_'}:}] omit the \forminline{node_} functions, which describe the fields and momenta which flow into each vertex.
    \item[\textbf{\texttt{\color{ptvcyan}`WithBlocks_'}:}] tag each graph's ``blocks'', sub-graphs which are connected to the rest of the graph by a single vertex, in the \forminline{block_} function.
    \item[\textbf{\texttt{\color{ptvcyan}`WithOnePISets_'}:}] tag each graph's one-particle irreducible subsets of vertices in the \forminline{onepi_} function.
\end{description}

With the above options and example models in mind, we now display some example output from the generator. The options:
\begin{formblock}
Local gluglu1 = diagrams_(QCD, {glu}, {glu}, int, ext, 1,
    `WithEdges_'+`OnShell_'+`NoTadpole_'+`NoSnail_');
\end{formblock}
produce:

\begin{minipage}{\textwidth} % force no page breaks in the listing
\begin{formblocksmallnohl}
gluglu1 =
 - topo_(1) * node_(1,1,glu(-p1)) * node_(2,1,glu(-p2))
  * node_(3,g,QUA(-k1),qua(-k2),glu(p1)) * node_(4,g,QUA(k2),qua(k1),glu(p2))
  * edge_(1,glu(p1),1,3) * edge_(2,glu(p2),2,4)
  * edge_(3,qua(k1),3,4) * edge_(4,QUA(k2),3,4)

 - topo_(1) * node_(1,1,glu(-p1)) * node_(2,1,glu(-p2))
  * node_(3,g,GHO(-k1),gho(-k2),glu(p1)) * node_(4,g,GHO(k2),gho(k1),glu(p2))
  * edge_(1,glu(p1),1,3) * edge_(2,glu(p2),2,4)
  * edge_(3,gho(k1),3,4) * edge_(4,GHO(k2),3,4)

 + 1/2 * topo_(1) * node_(1,1,glu(-p1)) * node_(2,1,glu(-p2))
  * node_(3,g,glu(p1),glu(-k1),glu(-k2)) * node_(4,g,glu(p2),glu(k1),glu(k2))
  * edge_(1,glu(p1),1,3) * edge_(2,glu(p2),2,4)
  * edge_(3,glu(k1),3,4) * edge_(4,glu(k2),3,4);
\end{formblocksmallnohl}
\end{minipage}

The \forminline{node\_} function arguments give their id number, the coupling
associated with the vertex they represent, and the \forminline{Particle} fields
which connect to them (which are functions of the incoming momenta).
The \forminline{edge\_} function arguments give their id number, the
\forminline{Particle} and momentum of the associated propagator, and the id
numbers of the \forminline{node\_} functions which they connect.
External particles have a special \forminline{node\_} function containing a
single field and a coupling of 1.

Specifying additionally \forminline{`TopologiesOnly_'} produces just the one
contributing topology:

\begin{minipage}{\textwidth} % force no page breaks in the listing
\begin{formblocksmallnohl}
+ topo_(1) * node_(1,1,-p1) * node_(2,1,-p2)
  * node_(3,g,p1,-k1,-k2) * node_(4,g,p2,k1,k2)
  * edge_(1,p1,1,3) * edge_(2,p2,2,4)
  * edge_(3,k1,3,4) * edge_(4,k2,3,4)
\end{formblocksmallnohl}
\end{minipage}

In this case, \forminline{Particle} information is omitted from the \forminline{node\_} and \forminline{edge\_} functions.
The \forminline{topo_} tags are consistent with the tags present in the full graph
output, produced when \forminline{`TopologiesOnly_'} is not specified.

To assist with debugging configuration problems or to see more information on
the internals of the diagram generator, one may specify \forminline{On GrccVerbose;}.

%%%%-----------------------------------------------------------------
\subsection{Floating-Point Coefficients}
\label{sec:float}
The implementation of arbitrary-precision floating point numbers had been planned since the earliest stages of the development of \FORM, but somehow there was never the immediate need. With \FORMfi{}, this capability is now fully available. 
Internally, \FORM{} relies on the low-level routines of the GMP library~\cite{Granlund12} for arbitrary floating point arithmetic and the MPFR library~\cite{10.1145/1236463.1236468,mpfr} for the numerical evaluation of the standard mathematical functions and constants. 

Floating point numbers are enabled via the preprocessor instruction \forminline{#StartFloat}. The following program illustrates this.

\begin{minipage}{\textwidth} % force no page breaks in the listing
\begin{formblock}
    #StartFloat 11d
    Local F = 103993/33102;
    ToFloat;
    Print "<1> %t";
    ToRational;
    Print "<2> %t";
    .end
<1>  + 3.141592653e+00
<2>  + 103993/33102
\end{formblock}
\end{minipage}

Here, the argument ``\texttt{11d}'' of \forminline{#StartFloat} specifies the working precision in decimal digits, as indicated by the \texttt{d}. One may also specify the precision in bits, using \texttt{b}, which corresponds to the internal precision used by GMP and MPFR\@. 
The example also shows the new \forminline{ToFloat} and \forminline{ToRational} statements, which convert rational coefficients to floating point coefficient and vice versa. Note that trailing zeroes are truncated in the floating-point output. The conversion in \forminline{ToRational} is implemented by means of continued fractions. 

A second instance of the \forminline{#StartFloat} is shown below:

\begin{minipage}{\textwidth} % force no page breaks in the listing
\begin{formblock}
    Symbols a,b;
    #StartFloat 10d, MZV=3
    Local F = a*mzv_(2,1)+b*euler_(-1,2);
    Evaluate mzv_ euler_;
    Print;
    .end
   F =
      2.695764795e-01*b + 1.202056903e+00*a; 
\end{formblock}
\end{minipage}

The second (optional) argument in \forminline{#StartFloat} is required when working with multiple zeta values (\forminline{mzv\_}) or Euler sums (\forminline{euler\_}); it specifies the maximum weight that may occur. The evaluation of these sums requires auxiliary tables whose size depends on this value. The default value is zero. 
The numerical evaluation itself is triggered by the \forminline{Evaluate} statement.  
The algorithm used for evaluating \forminline{mzv\_} and \forminline{euler\_} is based on~\cite{borwein1999specialvaluesmultiplepolylogarithms}.
In addition, \FORMfi{} provides the function \forminline{mzvhalf\_} for the numerical evaluation of the harmonic polylogarithms of argument $1/2$.
If one tries to evaluate a sum with a weight that exceeds the maximum weight, the program terminates with an error.

\forminline{Evaluate} can also be used to numerically evaluate a range of other functions. Since the early days of \FORM{}, the function names \forminline{sqrt\_}, \forminline{ln\_}, \forminline{sin\_}, \forminline{cos\_}, \forminline{tan\_}, \forminline{asin\_}, \forminline{acos\_}, \forminline{atan\_}, \forminline{atan2\_}, \forminline{sinh\_}, \forminline{cosh\_}, \forminline{tanh\_}, \forminline{asinh\_}, \forminline{acosh\_}, \forminline{atanh\_}, \forminline{li2\_}, \forminline{lin\_} have been reserved. 
All except \forminline{lin\_}, are part of the MPFR library and their numerical evaluation within \FORM{} uses MPFR accordingly. 
In particular, if \forminline{Evaluate} is given without arguments, all supported functions are evaluated.
Currently, numerical evaluation is limited to real arguments. This restriction may be lifted in the future if numerical evaluation over complex numbers is implemented.
Besides the functions, \forminline{Evaluate} can also be used for the numerical evaluation of the symbol \forminline{pi\_}, and the new reserved symbols \forminline{ee\_} and \forminline{em\_} for the mathematical constants $e$ and $\gamma_E$ respectively.

The list of reserved function names is further extended with \forminline{mzv\_}, \forminline{euler\_}, \forminline{mzvhalf\_}, \forminline{agm\_}, \forminline{gamma\_} and \forminline{eexp\_}. These, as well as the new \forminline{float\_} function discussed below, are reserved even if \FORM{} is compiled without floating point support.

Internally, floating point numbers are represented by the function \forminline{float\_}, whose arguments encode the internal representation of the floating point number. In a normalized term containing \forminline{float\_}, the rational coefficient will be either $1/1$ or $-1/1$, where the sign of the term is absorbed into the rational coefficient. 
Furthermore, \forminline{float\_} is protected from the pattern matcher and from statements that act on functions, such as \forminline{Transform}, \forminline{Argument}, \forminline{Normalize} etc.
The following program illustrates this:

\begin{minipage}{\textwidth} % force no page breaks in the listing
\begin{formblock}
    CFunction f;
    #StartFloat 10d
    Local F = 1.23456789 + f(1,2);
    Identify f?(?a) = f(10);
    Print "<1> %t";
    .sort
<1>  + 1.23456789e+00
<1>  + f(10)
    #EndFloat
    Normalize;
    Print "<2> %t";
    .sort
<2>  + float_(2,3,1,420101683733788795657820481376616399786)
<2>  + 10*f(1)
    #StartFloat 5d
    Print "<3> %t";
    .end
<3>  + 1.2346e+00
<3>  + 10*f(1)
\end{formblock}
\end{minipage}

As shown, the \forminline{Identify} statement does not effect the \forminline{float_} function. Here we also see the use of the preprocessor statement \forminline{\#EndFloat} which closes the floating point system. After this statement, the \forminline{float_} function becomes a regular function. Its protected status, however, persists so that \forminline{Identify} statements or statements like \forminline{Normalize} still do not modify it. 

Now that \FORM{} has floating point number arithmetic, there is need for statements that can do some basic operations. 
Besides the new statements described above to work with floating point numbers, we also introduce:
\begin{description}
\item[StrictRounding:] Rounds floating point numbers to a given precision. The syntax is \forminline{StrictRounding [precision];} where the optional precision uses the same syntax as in \forminline{\#StartFloat}. If omitted, the default precision given in \forminline{\#StartFloat} is used.
\item[Chop:] Removes floating point numbers that are smaller in absolute magnitude than a specified threshold. It takes one argument, delta: \forminline{Chop <delta>;}. All floating point numbers with magnitude less than delta are replaced by zero, while terms without floating point coefficient remain unaffected.
\end{description}
For further details on these and all other floating point features, we refer the reader to the manual.

%%%%-----------------------------------------------------------------
\subsection{\textsc{\textmd{Flint}} Interface for Polynomial Arithmetic}
\label{sec:flint}
\FORMfi{} contains an interface to the Fast Library for Number Theory (\FLINT)~\cite{flint}, which is used for low-level polynomial arithmetic operations. \FLINT{} version 3.2 or newer is required\footnote{
The required version may be increased in the future, if bugs in \FLINT{} are discovered which can't be worked around in the \FORM{} codebase.
}; the path to the library may be specified at \FORM{} compile time with the \texttt{configure} option \texttt{--with-flint=/path/to/flint}, in analogy to \FORM's build configuration for other external libraries such as GMP\@.

The following \FORM{} operations now make use of \FLINT{} routines internally: \forminline{PolyRatFun}, \forminline{FactArg}, \forminline{FactDollar}, \forminline{div\_}, \forminline{rem\_}, \forminline{mul\_}, \forminline{gcd\_} and \forminline{inverse\_}. Full expression factorization (through the use of \forminline{Factorize} and the \forminline{LocalFactorized} definition) and the \forminline{Modulus} modular-arithmetic mode do not yet use \FLINT{} routines, but support will be added in the future.

Apart from ensuring \FORM{} is correctly linked against \FLINT{} at compilation, nothing is required from the user at the level of their \FORM{} scripts; the syntax is identical\footnote{
Note that in some cases, the overall sign of the result of \forminline{gcd\_} may differ w.r.t.~previous versions.}
and the use of \FLINT{} is enabled by default. The use of \FLINT{} may be disabled and enabled on a per-module basis by specifying \forminline{Off flint;} and \forminline{On flint;} respectively.

\FORM{} scripts which make heavy use of these polynomial operations should find a large improvement in performance, particularly in multivariate cases. Table~\ref{tab:flinttimes} shows run times with and without \FLINT{} for a selection of benchmarks:
\begin{description}[leftmargin=!,labelwidth=\widthof{\texttt{minceex}:}]
    \item[\texttt{minceex}:] Compute the $N$th moment of a three-loop forward-scattering graph in deep-inelastic scattering. This is a univariate test, where term coefficients are rational polynomials of the dimensional regulator $\epsilon$. This is a slightly modified version of the example distributed in the ``Mincer exact'' package \url{https://www.nikhef.nl/~form/maindir/packages/mincer/mincerex.tgz}
    which is an $\epsilon$-exact version of the \textsc{Mincer} package~\cite{Gorishnii:1989gt,Larin:1991fz}.
    \item[\texttt{forcer}:] Integration-by-parts reduction of four-loop propagator integrals using the \textsc{Forcer} package~\cite{Ruijl:2017cxj}. Here we show numbers for integrals with 4, 5 and 6 ``dots'' on propagators. This is a univariate test, where term coefficients are rational polynomials of the dimensional regulator $\epsilon$.
    \item[\texttt{mbox1l}:] Integration-by-parts reduction of a one-loop four-point integral with massive internal propagators and two (equally) massive external legs. This is a multivariate test, where term coefficients are rational polynomials depending on $\epsilon,q_{12},q_{13},q_{33},m^2$ where $q_{ij}$ represents the scalar product of $q_i$ and $q_j$ and $m$ is the propagator mass. Reduction rules are computed using the \texttt{LiteRed}~\cite{Lee:2013mka} package and converted to \FORM{} notation. The ``Config.''~indices correspond to the propagator powers of the integral which is reduced.
\end{description}

\begin{table}[h]
    \centering
    \begin{tabular}{ll|rrr}
        \textbf{Benchmark} & \textbf{Config.} & \forminline{Off flint;} & \forminline{On flint;} & Speed-up \\
        \hline
        \texttt{minceex} & $N=8$ & 42.4 & 35.9 & 1.2x \\
                         & $N=10$ & 152 & 129 & 1.2x \\
                         & $N=12$ & 556 & 475 & 1.2x \\
        \hline
        \texttt{forcer} & 15-prop & 84.9 & 49.2 & 1.7x \\
                        & 16-prop & 232 & 123 & 1.9x \\
                        & 17-prop & 641 & 308 & 2.1x \\
        \hline
        \texttt{mbox1l} & (3,2,2,2) & 20.7 & 1.32 & 16x \\
                        & (3,3,2,2) & 85.2 & 2.9 & 29x \\
                        & (3,3,3,3) & 588 & 11.2 & 53x \\
    \end{tabular}
    \caption{Benchmark (wall) times, in seconds, comparing \FORM{} without and with the use of \FLINT's low-level polynomial routines. The tests were run on an AMD Ryzen 9 7900X with 24 \TFORM{} workers, using \FLINT{} v3.2.1.}
    \label{tab:flinttimes}
\end{table}

%%%%-----------------------------------------------------------------
\subsection{Sort-Buffer Reallocation}
It is well known that \FORM{} makes large memory allocations that may only be used for a small percentage
of its runtime. With \FORMfi{}, it is possible for the user to trigger
the reallocation of the small and large sorting buffers, which has the effect of reducing memory usage
as measured by ``resident set size'', if not ``virtual memory size'', and can reduce unnecessary swapping
of memory pages to disk by the operating system. This can be of particular benefit in \FORM{} scripts
which feature some very heavy modules with large expression swell, followed by a long series of light modules.

The reallocation can be triggered once, at the start of the current module, with the pre-processor directive
\begin{formblock}
#SortReallocate
\end{formblock}
or at the start of every module with
\begin{formblock}
On SortReallocate;
\end{formblock}
The second variant in particular may result in a noticeable performance impact, although this may be
recovered by the ability set larger buffer sizes. This depends entirely on what the user script does.

%%%%-----------------------------------------------------------------
\subsection{Sort file compression with Zstandard}
\FORMfi{} supports the use of the Zstandard compression library \cite{rfc8878}, in addition to the previous support for zlib~\cite{zlib}. Compression is used for the sort files, which store on-disk patches during a large sorting operation. It is enabled by default if the Zstandard library is found when compiling \FORM. It is enabled explicitly using
\begin{formblock}
On compress zstd;
\end{formblock}
and the use of zlib may still be forced using
\begin{formblock}
On compress gzip;
\end{formblock}
The improved performance of Zstandard with respect to zlib results in a runtime improvement of \FORM{} in sorting-dominated benchmarks by up to around 8\%, and the sort files are around 6\% smaller. Zstandard is not used for compression in tablebases, to maintain compatibility of tablebase files with older versions of \FORM.

%%%%-----------------------------------------------------------------
\subsection{Read-only TableBases}
Tablebases can now be opened in a read-only mode, meaning write access to the tablebase file is not required. This feature is useful, for example, when providing access to tablebase files to collaborators on a shared filesystem or to ensure that no additional entries can be accidentally added. The syntax is
\begin{formblock}
TableBase "filename.tbl" open readonly;
\end{formblock}
Attempting to add an entry to the table results in the error
\begin{formblocknohl}
Tablebase with the name filename.tbl opened in read only mode
\end{formblocknohl}
and terminates \FORM. Attempting to open a non-existing file in read-only mode results in the error
\begin{formblocknohl}
Trying to open non-existent TableBase in readonly mode: filename.tbl
\end{formblocknohl}
and terminates \FORM, rather than creating the tablebase as in the default read-write mode.

%%%%-----------------------------------------------------------------
\subsection{Backtracing}
On Linux systems, when compiled with the configure option \texttt{--enable-backtrace} and the utilities \texttt{eu-addr2line} or \texttt{addr2line} are available, \FORMfi{} will print a stack trace when terminating with an error condition to assist with bug reporting and debugging efforts. Due to changes in compilation options, this feature incurs a 1\% performance penalty in some (but not most!) benchmarks; for this reason it is disabled by default.

Nonetheless, it is strongly recommended to enable this feature to assist in debugging efforts. When enabled, a crash log will look something like the following,

\begin{minipage}{\textwidth} % force no page breaks in the listing
\begin{formblocknohl}
Program terminating at gcd-simple.frm Line 10 -->
Terminate called from polywrap.cc:156 (poly_gcd)
Backtrace:
# 0: TerminateImpl at startup.c:1870:10
# 1: poly_gcd at polywrap.cc:158:32
# 2: GCDfunction3 at ratio.c:1205:2
# 3: GCDfunction at ratio.c:1061:6
# 4: Generator at proces.c:4012:9
# 5: CatchDollar at dollar.c:112:6
# 6: PreProcessor at pre.c:1129:26
# 7: main at startup.c:1746:2
\end{formblocknohl}
\end{minipage}

and this output should be included in any bug reports raised against \FORM.
Note that backtrace output is also given when \FORM{} terminates due to, for example,
syntax errors in scripts; thus the appearance of a backtrace does not necessarily
imply the existence of a bug which should be reported.

%%%%%%%%%%%%%%%%%%%%%%%%%%%%%%%%%%%%%%%%%%%%%%%%%%%%%%%%%%%%%%%%%%%%%%
\section{Deprecations}
\label{sec:deprecations}
In this section we describe the deprecation of several features which,
to our knowledge, are not used. They contain known bugs and present a
substantial maintenance burden. In \FORMfi, use of these features will print a
deprecation warning, which can be suppressed by providing the command-line
option \texttt{-ignore-deprecation} or defining the environment variable
\texttt{FORM_IGNORE_DEPRECATION=1}.

If you use these features and would like to see continued support and
maintenance efforts, please follow the links given in the following
sections and provide some information regarding your use case.

\subsection{Native Windows Support}
\url{https://github.com/form-dev/form/issues/623}

We are not aware of anyone making use of native Windows builds of \FORM{} for
any serious computation. The existence of ``Windows Subsystem for Linux'' (WSL)
provides a practical alternative, making Windows support less necessary than
it might have been in the past. \FORM{} is mainly developed in Linux environments,
making maintenance of Windows builds rather difficult.

\subsection{32-bit System Support}
\url{https://github.com/form-dev/form/issues/624}

Now that 64-bit computing platforms are very much the norm, spending developer effort
on the maintenance of 32-bit support provides little benefit. The memory requirements
of cutting-edge computations means that 32-bit platforms are, to a large extent, limited
to solving ``toy problems''.

\subsection{ParFORM}
\url{https://github.com/form-dev/form/issues/625}

\PARFORM{}~\cite{Tentyukov:2002nwj} was developed at a time when computers had just a few
CPU cores and a few
10s of GB of memory, meaning that making use of the memory across multiple systems
was beneficial. Now that single systems may have several hundred CPU cores and multiple
TB of memory, the parallelisation methods of \TFORM{} on single systems are far superior.
The internal structure of \PARFORM{} is very different, making it difficult to effectively
maintain both.

\subsection{Checkpoint Mechanism}
\url{https://github.com/form-dev/form/issues/626}

Introduced in \FORM{} 4.0, the checkpoint mechanism enables the saving of intermediate
states of computation, enabling recovery after crashes or termination. Maintaining this
functionality requires significant effort and thorough testing; the current implementation
is very likely to contain bugs. We are not aware of any users who actively use this
mechanism.

%%%%%%%%%%%%%%%%%%%%%%%%%%%%%%%%%%%%%%%%%%%%%%%%%%%%%%%%%%%%%%%%%%%%%%
\section{Conclusions}
In this paper, we have demonstrated the major new features of \FORMfi{},
which substantially extend its capabilities for solving large-scale computer
algebra problems and reduce dependence on non-free software.

There are many avenues through which \FORM{} can be further developed in the
future, to support the requirements of its user base. To that end, we strongly
encourage any user with feature ideas to propose them here
\url{https://github.com/form-dev/form/discussions} and to attend the
annual \FORM{} Developers' Workshop series. Bug reports and interesting
test cases are also greatly appreciated and contribute to the robustness
of the software.

Following the retirement of J.~Vermaseren in 2018, who has been the primary developer of \FORM{} since its inception, the continued evolution of the system increasingly depends on broader support and community involvement. A sustained, collaborative development effort will be essential to ensure that \FORM{} remains a reliable and high-performance computer algebra system for the specialised needs of the theoretical high-energy physics community.

%%%%%%%%%%%%%%%%%%%%%%%%%%%%%%%%%%%%%%%%%%%%%%%%%%%%%%%%%%%%%%%%%%%%%%
\section*{Acknowledgements}

The work of J.~D.~was supported by STFC Consolidated Grant ST/X000699/1.
The work of T.~U.~was supported by JSPS KAKENHI Grant Numbers JP22K03604 and JP24K07055.

%%%%%%%%%%%%%%%%%%%%%%%%%%%%%%%%%%%%%%%%%%%%%%%%%%%%%%%%%%%%%%%%%%%%%%

\bibliographystyle{elsarticle-num}
\bibliography{main.bib}

@article{Vermaseren:2000nd,
    author = "Vermaseren, J. A. M.",
    title = "{New features of FORM}",
    eprint = "math-ph/0010025",
    archivePrefix = "arXiv",
    month = "10",
    year = "2000"
}

@article{Kuipers:2012rf,
    author = "Kuipers, J. and Ueda, T. and Vermaseren, J. A. M. and Vollinga, J.",
    title = "{FORM version 4.0}",
    eprint = "1203.6543",
    archivePrefix = "arXiv",
    primaryClass = "cs.SC",
    reportNumber = "NIKHEF-2012-004, TTP12-008, SFB-CPP-12-15",
    doi = "10.1016/j.cpc.2012.12.028",
    journal = "Comput. Phys. Commun.",
    volume = "184",
    pages = "1453--1467",
    year = "2013"
}

@article{Ruijl:2017dtg,
    author = "Ruijl, Ben and Ueda, Takahiro and Vermaseren, Jos",
    title = "{FORM version 4.2}",
    eprint = "1707.06453",
    archivePrefix = "arXiv",
    primaryClass = "hep-ph",
    month = "7",
    year = "2017"
}

@misc{borwein1999specialvaluesmultiplepolylogarithms,
      title={Special Values of Multiple Polylogarithms}, 
      author={Jonathan M. Borwein and David M. Bradley and David J. Broadhurst and Petr Lisonek},
      eprint={math/9910045},
      archivePrefix={arXiv},
      primaryClass={math.CA},
      journal={Transactions of the American Mathematical Society},
      volume={353},
      number={3},
      pages={907--941},
      year={2001}
}

@Manual{Granlund12,
  title = 	 "{GNU MP}: {T}he {GNU} {M}ultiple {P}recision
		  {A}rithmetic {L}ibrary",
  author = "Torbjörn Granlund and {the GMP development team}",
  year = 2023,
  note = "{Version 6.3.0, \url{http://gmplib.org/}}"
}

@Manual{zlib,
  title = "zlib",
  author = "J Gailly and M. Adler",
  year = 2025,
  note = "{Version 1.3.1, \url{https://zlib.net/}}"
}

@article{10.1145/1236463.1236468,
author = {Fousse, Laurent and Hanrot, Guillaume and Lef\`{e}vre, Vincent and P\'{e}lissier, Patrick and Zimmermann, Paul},
title = {MPFR: A multiple-precision binary floating-point library with correct rounding},
year = {2007},
issue_date = {June 2007},
publisher = {Association for Computing Machinery},
address = {New York, NY, USA},
volume = {33},
number = {2},
issn = {0098-3500},
url = {https://doi.org/10.1145/1236463.1236468},
doi = {10.1145/1236463.1236468},
journal = {ACM Trans. Math. Softw.},
month = jun,
pages = {13–es},
numpages = {15}
}

@manual{mpfr,
   title   = "{GNU MPFR}: {T}he {M}ultiple {P}recision {F}loating-{P}oint {R}eliable {L}ibrary",
  note     = "{Version 4.2.2, \url{https://www.mpfr.org/}}",
  year = 2025
}

@manual{flint,
  key = {{FLINT}},
  author = {{The {FLINT} team}},
  title = {{FLINT}: {F}ast {L}ibrary for {N}umber {T}heory},
  year = {2025},
  note = {Version 3.2.0, \url{https://flintlib.org}}
}

@article{Kaneko:1994fd,
    author = "Kaneko, Toshiaki",
    title = "{A Feynman graph generator for any order of coupling constants}",
    eprint = "hep-th/9408107",
    archivePrefix = "arXiv",
    reportNumber = "MGU-CS-94-01, KEK-CP-020, KEK-PREPRINT-94-83",
    doi = "10.1016/0010-4655(95)00122-6",
    journal = "Comput. Phys. Commun.",
    volume = "92",
    pages = "127--152",
    year = "1995"
}

@article{Nogueira:1991ex,
    author = "Nogueira, Paulo",
    title = "{Automatic Feynman Graph Generation}",
    reportNumber = "IFM-7-91",
    doi = "10.1006/jcph.1993.1074",
    journal = "J. Comput. Phys.",
    volume = "105",
    pages = "279--289",
    year = "1993"
}

@article{Tentyukov:2007mu,
    author = "Tentyukov, M. and Vermaseren, J. A. M.",
    title = "{The Multithreaded version of FORM}",
    eprint = "hep-ph/0702279",
    archivePrefix = "arXiv",
    reportNumber = "NIKHEF-07-005, SFB-CPP-07-08, TTP07-06",
    doi = "10.1016/j.cpc.2010.04.009",
    journal = "Comput. Phys. Commun.",
    volume = "181",
    pages = "1419--1427",
    year = "2010"
}

@article{Kuipers:2013pba,
    author = "Kuipers, J. and Ueda, T. and Vermaseren, J. A. M.",
    title = "{Code Optimization in FORM}",
    eprint = "1310.7007",
    archivePrefix = "arXiv",
    primaryClass = "cs.SC",
    reportNumber = "NIKHEF-2013-036, TTP13-031, SFB-CPP-13-80",
    doi = "10.1016/j.cpc.2014.08.008",
    journal = "Comput. Phys. Commun.",
    volume = "189",
    pages = "1--19",
    year = "2015"
}

@article{Ueda:2020wqk,
    author = "Ueda, T. and Kaneko, T. and Ruijl, B. and Vermaseren, J. A. M.",
    title = "{Further developments of FORM}",
    doi = "10.1088/1742-6596/1525/1/012013",
    journal = "J. Phys. Conf. Ser.",
    volume = "1525",
    pages = "012013",
    year = "2020"
}

@article{Ruijl:2017cxj,
    author = "Ruijl, B. and Ueda, T. and Vermaseren, J. A. M.",
    title = "{Forcer, a FORM program for the parametric reduction of four-loop massless propagator diagrams}",
    eprint = "1704.06650",
    archivePrefix = "arXiv",
    primaryClass = "hep-ph",
    reportNumber = "NIKHEF-2017-019",
    doi = "10.1016/j.cpc.2020.107198",
    journal = "Comput. Phys. Commun.",
    volume = "253",
    pages = "107198",
    year = "2020"
}

@article{Lee:2013mka,
    author = "Lee, Roman N.",
    editor = "Wang, Jianxiong",
    title = "{LiteRed 1.4: a powerful tool for reduction of multiloop integrals}",
    eprint = "1310.1145",
    archivePrefix = "arXiv",
    primaryClass = "hep-ph",
    doi = "10.1088/1742-6596/523/1/012059",
    journal = "J. Phys. Conf. Ser.",
    volume = "523",
    pages = "012059",
    year = "2014"
}

@article{Kardos:2025klp,
    author = "Kardos, Adam and Moch, Sven-Olaf and Schnetz, Oliver",
    title = "{HyperFORM -- a FORM package for parametric integration with hyperlogarithms}",
    eprint = "2511.19992",
    archivePrefix = "arXiv",
    primaryClass = "hep-ph",
    month = "11",
    year = "2025"
}

@article{Duhr:2025oil,
    author = "Duhr, Claude and Mukherjee, Pooja and Vasquez, Andres",
    title = "{ANATAR: AN Automated Tool for higher-order Amplitude geneRation}",
    eprint = "2509.13951",
    archivePrefix = "arXiv",
    primaryClass = "hep-ph",
    reportNumber = "DESY-25-117, BONN-TH-2025-28",
    month = "9",
    year = "2025"
}

@article{Brod:2024zaz,
    author = {Brod, Joachim and H{\"u}depohl, Lorenz and Stamou, Emmanuel and Steudtner, Tom},
    title = "{MaRTIn -- Manual for the ``Massive Recursive Tensor Integration''}",
    eprint = "2401.04033",
    archivePrefix = "arXiv",
    primaryClass = "hep-ph",
    doi = "10.1016/j.cpc.2024.109372",
    journal = "Comput. Phys. Commun.",
    volume = "306",
    pages = "109372",
    year = "2025"
}

@article{Goode:2024cfy,
    author = "Goode, Jae and Herzog, Franz and Teale, Sam",
    title = "{OPITeR: A program for tensor reduction of multi-loop Feynman integrals}",
    eprint = "2411.02233",
    archivePrefix = "arXiv",
    primaryClass = "hep-ph",
    doi = "10.1016/j.cpc.2025.109606",
    journal = "Comput. Phys. Commun.",
    volume = "312",
    pages = "109606",
    year = "2025"
}

@article{Heinrich:2023til,
    author = "Heinrich, G. and Jones, S. P. and Kerner, M. and Magerya, V. and Olsson, A. and Schlenk, J.",
    title = "{Numerical scattering amplitudes with pySecDec}",
    eprint = "2305.19768",
    archivePrefix = "arXiv",
    primaryClass = "hep-ph",
    reportNumber = "KA-TP-09-2023, P3H-23-035, IPPP/23/24",
    doi = "10.1016/j.cpc.2023.108956",
    journal = "Comput. Phys. Commun.",
    volume = "295",
    pages = "108956",
    year = "2024"
}

@article{Tentyukov:2002nwj,
    author = "Tentyukov, M. and Fliegner, D. and Frank, M. and Onischenko, A. and Retey, A. and Staudenmaier, H. M. and Vermaseren, J. A. M.",
    editor = "Bhat, P. C. and Kasemann, M.",
    title = "{ParFORM: Parallel version of the symbolic manipulation program FORM}",
    eprint = "cs/0407066",
    archivePrefix = "arXiv",
    reportNumber = "SFB-CPP-04-27, TTP04-15",
    doi = "10.1063/1.1405304",
    journal = "AIP Conf. Proc.",
    volume = "583",
    number = "1",
    pages = "202",
    year = "2002"
}

@article{vanHoegaerden:2025nmh,
    author = "van Hoegaerden, Paul A. J. W. and Marinissen, Coenraad B. and Waalewijn, Wouter J.",
    title = "{IBIS: Inverse BInomial sum Solver}",
    eprint = "2506.19904",
    archivePrefix = "arXiv",
    primaryClass = "hep-ph",
    month = "6",
    year = "2025"
}

@article{Braun:2025afl,
    author = {Braun, Jens and Campillo Aveleira, Benjamin and Heinrich, Gudrun and H{\"o}fer, Marius and Jones, Stephen P. and Kerner, Matthias and Lang, Jannis and Magerya, Vitaly},
    title = "{One-loop calculations in effective field theories with GoSam-3.0}",
    eprint = "2507.23549",
    archivePrefix = "arXiv",
    primaryClass = "hep-ph",
    reportNumber = "KA-TP-21-2025, P3H-25-051, IPPP/25/50, CERN-TH-2025-134",
    doi = "10.21468/SciPostPhysCodeb.62",
    journal = "SciPost Phys. Codeb.",
    volume = "62",
    pages = "1",
    year = "2026"
}

@article{Heller:2021qkz,
    author = "Heller, Matthias and von Manteuffel, Andreas",
    title = "{MultivariateApart: Generalized partial fractions}",
    eprint = "2101.08283",
    archivePrefix = "arXiv",
    primaryClass = "cs.SC",
    reportNumber = "MITP/21-002, MSUHEP-20-016",
    doi = "10.1016/j.cpc.2021.108174",
    journal = "Comput. Phys. Commun.",
    volume = "271",
    pages = "108174",
    year = "2022"
}

@article{Larin:1991fz,
    author = "Larin, S. A. and Tkachov, F. V. and Vermaseren, J. A. M.",
    title = "{The FORM version of MINCER}",
    reportNumber = "NIKHEF-H-91-18",
    month = "9",
    year = "1991"
}

@misc{rfc8878,
    series =    {Request for Comments},
    number =    8878,
    howpublished =  {RFC 8878},
    publisher = {RFC Editor},
    doi =       {10.17487/RFC8878},
    url =       {https://www.rfc-editor.org/info/rfc8878},
    author =    {Yann Collet and Murray Kucherawy},
    title =     {{Zstandard Compression and the `application/zstd' Media Type}},
    pagetotal = 45,
    year =      2021,
    month =     feb,
}

@article{Gorishnii:1989gt,
    author = "Gorishnii, S. G. and Larin, S. A. and Surguladze, L. R. and Tkachov, F. V.",
    title = "{Mincer: Program for Multiloop Calculations in Quantum Field Theory for the Schoonschip System}",
    doi = "10.1016/0010-4655(89)90134-3",
    journal = "Comput. Phys. Commun.",
    volume = "55",
    pages = "381--408",
    year = "1989"
}

@article{Feng:2021kha,
    author = "Feng, Feng and Xie, Yi-Fan and Zhou, Qiu-Chen and Tang, Shan-Rong",
    title = "{HepLib: A C++ library for high energy physics}",
    eprint = "2103.08507",
    archivePrefix = "arXiv",
    primaryClass = "hep-ph",
    doi = "10.1016/j.cpc.2021.107982",
    journal = "Comput. Phys. Commun.",
    volume = "265",
    pages = "107982",
    year = "2021"
}

@article{Shtabovenko:2025lxq,
    author = "Shtabovenko, Vladyslav",
    title = "{FeynCalc 10.2 and FeynHelpers 2: Multiloop calculations streamlined}",
    eprint = "2512.19858",
    archivePrefix = "arXiv",
    primaryClass = "hep-ph",
    reportNumber = "P3H-25-112, SI-HEP-2025-31",
    month = "12",
    year = "2025"
}

\end{document}